\documentclass[9pt,twocolumn,twoside]{pnas-new}

\usepackage{amsmath, bm}
\usepackage{graphicx}
\usepackage{color}
\usepackage{soul}
\usepackage{setspace}
\usepackage{multirow}
\usepackage{subfig}
\usepackage[title]{appendix}
\usepackage{rotating}
\usepackage{colortbl}
\definecolor{coloring}{rgb}{0.47, 0.53, 0.42}
\usepackage{chngcntr} 
\templatetype{pnasresearcharticle} 

\title{Detecting fake review buyers using network structure: Direct evidence from Amazon}

\author[a,1]{Sherry He}
\author[a,1,2]{Brett Hollenbeck} 
\author[b,1]{Gijs Overgoor}
\author[c,1]{Davide Proserpio}
\author[b,1]{Ali Tosyali} 

\affil[a]{Anderson School of Management, University of California, Los Angeles, Los Angeles, CA 90095}
\affil[b]{Saunders College of Business, Rochester Institute of Technology, Rochester, NY 14623}
\affil[c]{Marshall School of Business, University of Southern California, Los Angeles, CA 90089}

\leadauthor{He} 

\significancestatement{Online reviews significantly impact consumers decisions and are seen as crucial to the success of online markets. Despite this, the prevalence of fake reviews is arguably higher than ever, despite two decades of academic research on identifying and regulating them. We use novel data in which we directly observe which products buy fake reviews and study how to identify them. We show that products buying fake reviews are highly clustered in the product-reviewer network due to their reliance on common reviewers. This allows us to detect them with high accuracy using both supervised and unsupervised methods. Unlike approaches relying on reviews' text, this approach is more robust to manipulation by sellers. Moreover, it is both scalable and generalizable to many settings.}

\authorcontributions{Please provide details of author contributions here.}
\authordeclaration{The authors declare no conflict of interest.}
\equalauthors{\textsuperscript{1}Authors listed in alphabetical order}
\correspondingauthor{\textsuperscript{2}To whom correspondence should be addressed. E-mail: brett.hollenbeck@anderson.ucla.edu}


\begin{abstract}
Online reviews significantly impact consumers' decision-making process and firms' economic outcomes and are widely seen as crucial to the success of online markets. Firms, therefore, have a strong incentive to manipulate ratings using fake reviews. This presents a problem that academic researchers have tried to solve over two decades and on which platforms expend a large amount of resources. Nevertheless, the prevalence of fake reviews is arguably higher than ever. To combat this, we collect a dataset of reviews for thousands of Amazon products and develop a general and highly accurate method for detecting fake reviews. A unique difference between previous datasets and ours is that we directly observe which sellers buy fake reviews. Thus, while prior research has trained models using lab-generated reviews or proxies for fake reviews, we are able to train a model using actual fake reviews. We show that products that buy fake reviews are highly clustered in the product-reviewer network. Therefore, features constructed from this network are highly predictive of which products buy fake reviews. We show that our network-based approach is also successful at detecting fake reviews even without ground truth data, as unsupervised clustering methods can accurately identify fake review buyers by identifying clusters of products that are closely connected in the network. While text or metadata can be manipulated to evade detection, network-based features are more costly to manipulate because these features result directly from the inherent limitations of buying reviews from online review marketplaces, making our detection approach more robust to manipulation.
\end{abstract}

\dates{This manuscript was compiled on October 22, 2022}
\doi{\url{https://doi.org/10.1073/pnas.2211932119}}

\begin{document}

\maketitle
\thispagestyle{firststyle}
\ifthenelse{\boolean{shortarticle}}{\ifthenelse{\boolean{singlecolumn}}{\abscontentformatted}{\abscontent}}{}

\dropcap{O}nline reviews have a significant impact on consumer purchase decisions and are widely seen as crucial to the success of online markets~\cite{tadelis2016reputation,levi_trust}. Review and rating systems allow buyers and sellers to develop credible reputations in settings that are otherwise mostly anonymous. Because these reputations are crucial for seller outcomes, sellers have a large incentive to manipulate their ratings and inflate their reputation. As a result, online review platforms like Amazon, Yelp, or Tripadvisor have struggled since their inception with the problem of sellers manipulating their ratings with fake reviews. Rating manipulation can potentially cause buyers to buy from lower-quality sellers than they otherwise would, allow sellers to charge higher prices than if their true reputation was observed, and lower trust in reviews and review platforms, making it difficult for high-quality and honest sellers to compete. There is growing empirical evidence that fake reviews harm consumers~\cite{he2022market,Akesson_etal_welfare}. In addition to violating the platform's policies, these practices are the subject of ongoing investigations by the FTC, the UKCMA, and other regulators. 

Despite the vast amount of academic research over the past two decades (see~\cite{wu2020fake} for an extensive review) and the large amounts of time, effort, and money invested by online platforms to detect and remove fake reviews, they are nevertheless as prevalent as ever. Recent studies have found that millions of products on Amazon are using fake reviews, a large share of all reviews are fake~\cite{he2022market}, and consumers express very low levels of trust in online reviews as a result~\cite{dong_trust}.
    
Most research on the challenge of fake review detection relies on machine learning techniques that exploit features associated with reviews, such as ratings, helpful votes, and text content~\cite{wu2020fake}. A primary challenge in this approach is the lack of ground-truth data with which to train and test models. In other words, in order to develop models that can identify fake reviews, one must first have a corpus of existing fake reviews (and real reviews) with which to train the model. Researchers have attempted to overcome the inherent challenge this poses largely by using lab-generated reviews and considering platform-filtered reviews as a proxy for fake reviews. Scholars have criticized these approaches as having serious limitations~\cite{Heydari2015DetectionOR,Mukherjee2013WhatYF,Crawford2015SurveyOR,Vidanagama2019DeceptiveCR,wu2020fake}. Lab-generated fake reviews may lack authenticity, and bot-generated or other low-quality reviews (those from non-English speakers or containing many grammatical errors) at best pick the low-hanging fruit and miss the bulk of actual fake reviews. Moreover, using only reviews that are flagged and filtered by the platform's algorithms by definition cannot progress our understanding of how to identify the large number of fake reviews that currently evade those filters~\cite{he2022market}; and methods trained on these data have been repeatedly shown to lack external validity~\cite{Crawford2015SurveyOR}.

In addition to the lack of high-quality ground truth data, another shortcoming of current methods is that sophisticated actors can potentially evade filtering algorithms trained on historical features. For example, methods relying on text analysis are fundamentally limited in that, even with sophisticated models, human reviewers have strong incentives to evade detection and will therefore strive to write fake reviews that are indistinguishable from organic reviews. If particular phrases or tendencies become used to filter reviews, they can then avoid using these in their reviews. More generally, extant approaches to detecting which products are manipulating their ratings have not grappled with the economic incentives of buyers and sellers. We use unique data in which we observe sellers buy fake reviews and demonstrate how these economic incentives can be harnessed to design an approach that can detect fake review buyers with high accuracy and with limited scope for evasion.

We argue that these problems can be overcome by focusing on the product-reviewer network and identifying products that buy fake reviews rather than the fake reviews themselves. We hand-collect a large new dataset that provides accurate ground truth data by directly identifying a large and representative set of products that buy fake reviews on Amazon.com. Using this data, we study the relative effectiveness of different approaches for detecting what products manipulate their ratings. We find that, compared to products not using fake reviews, products that use fake reviews are highly connected and clustered in the network, implying that products using fake reviews tend to have more reviewers in common than other products. This follows naturally from the fact that, while regular products receive their reviews from a dispersed set of millions of Amazon customers, products buying fake reviews must rely on the relatively small number of reviewers participating in the fake review marketplace. Therefore, features derived from the product-reviewer network are especially useful for regulating fake reviews because, in addition to being more predictive than text or metadata features, they are more difficult or costly to manipulate by sellers than text or review timing, because these features result directly from the inherent limitations of acquiring reviews from fake review marketplaces.

\section*{Data}\label{sec:data}

To analyze fake review behavior on Amazon, we begin by collecting data from the private Facebook groups in which sellers buy reviews. While there are inherent limitations in observing fake review activity, these groups are generally seen as the primary channel by which sellers find reviewers.


From March 2020 to October 2020, we identify about 23 fake review-related groups daily. These groups are large and quite active, each having about 16,000 members on average and 568 fake review requests posted per day per group. Within these Facebook groups, sellers can obtain a five-star review that looks organic. Sellers post product pictures and review requests, after which the potential reviewer and the seller communicate via private Facebook messages. The vast majority of sellers that we observe buying fake reviews compensate the reviewer by refunding the cost of the product via a PayPal transaction after the five-star review has been posted along with the cost of the PayPal fee, sales tax, and in some cases, an additional commission. Reviewers are compensated for creating realistic seeming five-star reviews that evade Amazon's filters. This process differs from ``incentivized reviews,'' where sellers offer free or discounted products or discounts on future products in exchange for reviews that disclose the transaction and are not required to be five stars~\cite{burtch_hon_bapna_reviews}. 

Among these groups, and during our entire observation period, we observe Amazon sellers buying only positive reviews. As~\cite{he2022market} points out, this is likely because buying fake negative reviews to hurt competitors is costlier as the buyer needs to incur the full cost of the competitor's product. In addition, the benefit of purchasing fake negative reviews is likely to be lower than that of buying own positive reviews because the shift in own sales caused by a negative review on a competitor's page is indirect and dispersed across potentially many other competitor products besides the competitor for which the negative fake review is bought.


To identify which products are buying fake reviews, we hire a group of research assistants to visit these groups and select a random sample of products posted in them.
We collect data from these random Facebook fake review groups using this procedure on a weekly basis from October 2019 to June 2020, and the result is a sample of roughly 1,500 unique products. 

After identifying products whose ratings are manipulated, we collect data for these products on Amazon.com. We collect reviews and ratings for each of the products on a daily basis. For each review, we observe the rating, product ID, review text, review photos, and helpful votes. 

In addition to collecting this data for the products buying fake reviews, we collect daily review data for a set of 2,714 competitor products to serve as a comparison set. To do so, for each product buying fake reviews, we select the two products that appear most frequently in the search results during a period covering seven days before and after the date of the product's first Facebook post. The rationale is that we want to create a comparison set of products that are in the same subcategory as the products buying fake reviews and have a similar search rank before fake reviews are posted.

\section*{Identifying which products buy fake reviews}\label{sec:results}
There is an extensive literature on fake review detection on online platforms such as Amazon, Yelp, and Tripadvisor. This literature has proposed methods for detection based on text features, image features, spatiotemporal differences, network features, sentiment, and others (see the many references in~\cite{wu2020fake}). One of the main hurdles in creating algorithms capable of detecting fake reviews is the lack of ground truth data, i.e. reviews that are known to be fake, and this literature has been criticized for relying on low-quality proxies~\cite{wu2020fake}.


We overcome this issue by focusing on the products that buy fake reviews. As described in the Data Section, our dataset allows us to know with a high degree of certainty which Amazon products bought fake reviews, therefore providing accurate ground truth for our product-level analysis. We can then compare the performance of previously suggested methods and, in particular, test the performance of models that take advantage of the structure of the product-reviewer network. 
Another recent study~\citep{oak2021fault} has also utilized ground-truth data obtained by monitoring Facebook groups to perform review- and product-level fake review detection. However, the authors do not use the the network structure as we do in this paper. Despite this, their qualitative analysis and interviews performed with buyers and sellers align with several of our findings, and the authors conclude that the network structure of the products buying fake reviews could be useful to detect products buying fake reviews.


\subsection*{Network construction and network features generation}\label{sec:nw}
\begin{figure*}[!ht]
    \centering
    \includegraphics[width=.7\linewidth]{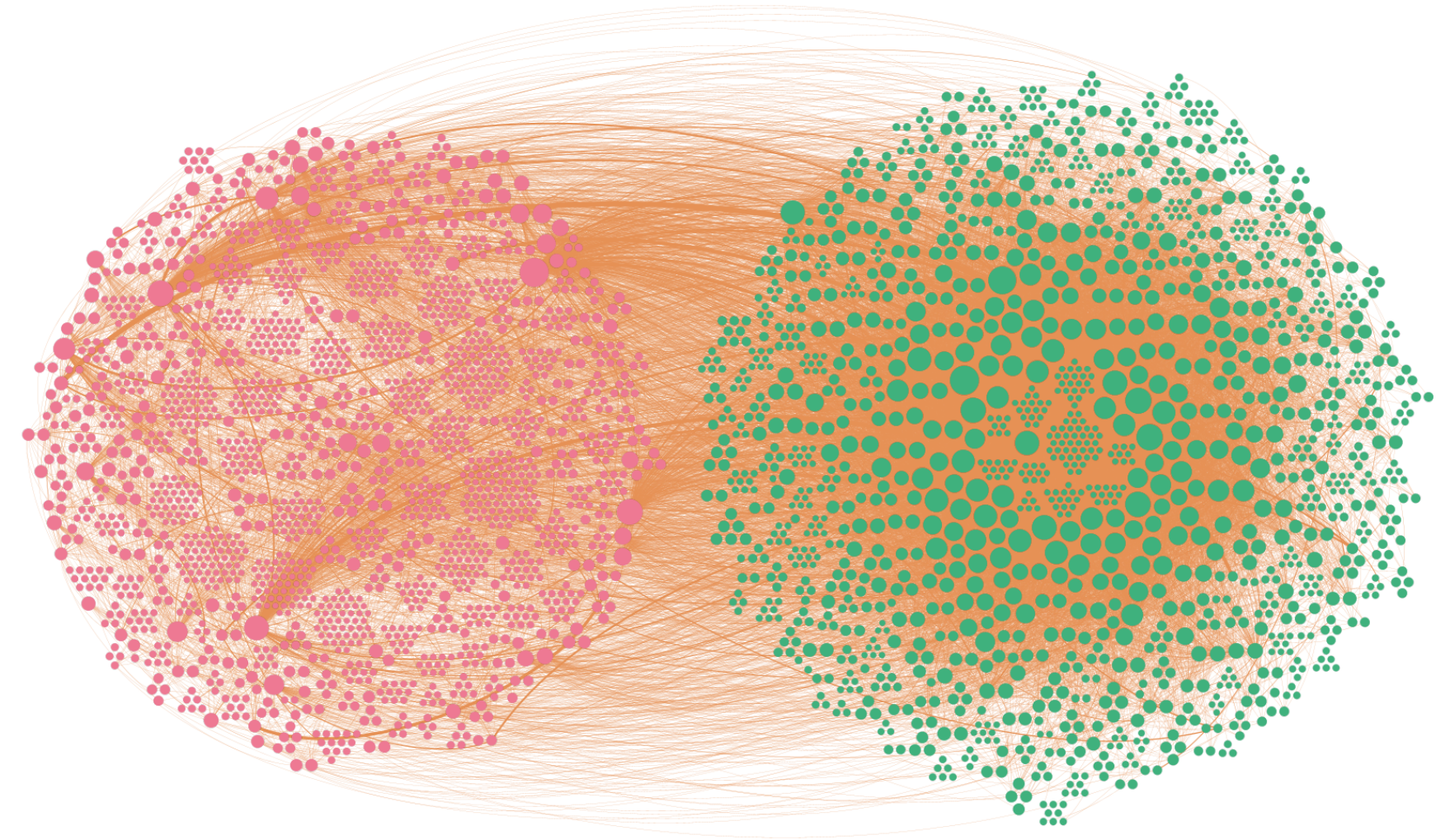}
    \caption{The network structure of Amazon products in our dataset. Green-colored nodes are the fake review buyers. We filter the edges to show only the ones between pairs of products with $\geq 2$ reviewers in common. A product’s size is proportional to the number of reviewers in common with other products. The figure shows that fake review buyers are mainly the larger nodes in the network and are highly clustered, highlighting the dense connectivity between fake review buyers due to their participation in the fake review marketplace.}
    \label{fig:network}
\end{figure*}

Our approach to detecting sellers buying fake reviews exploits the network structure of Amazon products.
We start by constructing a product network $G=(V, E)$ using our data, where $G$ is the network, $V$ is the set of nodes (i.e., products), and $E$ is the set of edges. An edge between two products represents the existence of common reviewers. Figure~\ref{fig:network} shows the network structure of Amazon products in our dataset. Then, for each node (i.e., product), we compute its degree, eigenvector centrality, PageRank score, and clustering coefficient. We provide the mathematical details of how we compute these measures in Section 1 of SI Appendix.

The degree of a product is the total number of reviewers it has in common with other products in the network. Eigenvector centrality measures the structural importance by considering the product's proximity to other structurally important products in the network. A product's centrality score increases by having reviewers in common with other products that themselves share many reviewers with other products in the network. PageRank, like eigenvector centrality, considers the importance of a product's neighbors when assigning a score to it. However, it mainly differs from eigenvector centrality by normalizing a neighbor product's structural importance by the number of its neighbors.

The above three measures (degree, eigenvector centrality, and PageRank) help us compare products' structural importance and understand how they relate to the overall network. In addition, we measure the products' connectivity within their neighborhood using the clustering coefficient. The clustering coefficient checks all pairs of a product's neighbors and considers how many of them have reviewers in common.

\subsection*{Additional features}
To compare the performance of different types of information at detecting fake review buyers, we also generate features from available product metadata and review content such as review text, ratings, timestamps, and images. 

The first set of features is generated from available metadata. These features include the number of reviews, average review rating, summary statistics of the time gap between reviews, the ratio of reviews that other consumers found helpful, the ratio of reviews with 1-star review rating, the ratio of reviews with 5-star review rating, the ratio of reviews that include review images, variation of review lengths for a product, and the average text similarity among the reviews of a product. 
These metadata should capture evidence of rating manipulation, such as having a disproportionate share of five-star reviews, excessive helpfulness votes, or odd timing characteristics such as long gaps in the arrival of reviews followed by the appearance of many at once.

We generate a second set of product-level features using review and product images. First, we use a pre-trained deep Convolutional Neural Network~\cite{he2016deep} for all images (both product and review) to extract the features as shown in Fig. S1 (SI Appendix). We then calculate the angular similarity of all the images belonging to the same product using the cosine distance between the vector representation of each image. Cosine distance is particularly effective for high-dimensional data~\cite{france2012distance}. We calculate three sets of features that summarize the degree of image similarity between all product reviews (controlling for multiple images belonging to the same review), the similarities between the seller's product images and the review images, and the similarity among all pairs of review images. For each set of features, we calculate the minimum, maximum, average, and standard deviation of the pair-wise similarities within a product. 

The third set of product-level features generated from the review content is text features. We first combine the review bodies of a product and treat each product as a document. Then, we calculate the TF-IDF score of words in each document. We only consider the top 1000 features that receive the highest TF-IDF scores for each product. Finally, each product is represented by a feature vector of length 1000.

Table~\ref{tab:features_desc} shows all the product-level features we generate to consider various aspects of reviews, products, and reviewer behaviors to detect fake review buyers. Table S1 (SI Appendix) shows the correlation coefficients between image, metadata, and network features.
    
\begin{table}[ht]
\centering
\small
\caption{Product-level features generated from review content and product network.}\label{tab:features_desc}
\resizebox{\linewidth}{!}{\begin{tabular}{c|l|l}
    \hline
    Type & Feature & Description\\
    \hline
    \multirow{4}{*}{\rotatebox[origin=c]{90}{Network}} 
     & degree & Total number of reviewers in common with other products\\
     & clustering coef & Clustering coefficient of the product\\
     & eigenvector cent & Eigenvector centrality score of the product\\
     & pagerank & PageRank score of the product\\
    \hline
    \multirow{8}{*}{\rotatebox[origin=c]{90}{Metadata}} 
     & tf-idf sim & Avg similarity of TF-IDF features b/w reviews of a product \\
     & \# reviews & Number of reviews\\
     & avg review rating & Avg review ratings \\
     & time b/w reviews & Avg, min, max, and stdev of time in days b/w reviews\\
     & share helpful & Share of reviews with helpful votes\\
     & share 1star & Share of reviews with 1-star rating\\
     & share 5star & Share of reviews with 5-star rating\\
     & share photo & Share of reviews with review photo\\
     & stdev review len & Variation of review lengths of a product\\
    \hline
    \multirow{3}{*}{\rotatebox[origin=c]{90}{Image}} 
    & img sim& Avg, min, max, and stdev of image similarity b/w product reviews \\
    & sim review & Avg, min, max, and stdev of similarity b/w all pairs of review images \\
    & sim product& Avg, min, max, and stdev of similarity b/w product and review images \\
    \hline
    \multirow{2}{*}{\rotatebox[origin=c]{90}{Text}} &\multirow{2}{*}{tf-idf} &\multirow{2}{*}{Top 1000 TF-IDF features of a product} \\
    & & \\
    \hline
\end{tabular}}
\end{table}
\section*{Results}
\subsection*{Supervised approach}
To test the predictive performance of the features we created, we employ a set of random forest classifiers.\footnote{We also test other classifiers and obtain qualitatively similar results, which can be found in Section 3 of SI Appendix.} Following the standard approach in the literature~\cite{zhang2022can,kumar2018detecting}, we train the classifiers on a random subset of 80\% of the products and evaluate its prediction performance on the remaining 20\%. We measure the area under the receiver operating characteristic curve (AUC), classification accuracy, true positive rate (TPR), true negative rate (TNR), and the F1 score. The technical details of our estimated model and model building process are in Section 2 of SI Appendix.


Table~\ref{tab:RF_pred_results} shows the performance of each type of feature set. We find that features constructed from the product-reviewer network outperform metadata, text, and image features across all accuracy metrics. The combined model that includes all features performs only slightly better than the model using only network features. In addition to performing best on balanced measures of accuracy, the network feature model performs best on the true positive rate. This is especially important for a rating platform for whom avoiding false positives is an important goal.

\begin{figure}[!h]
    \centering
    \includegraphics[width=\linewidth]{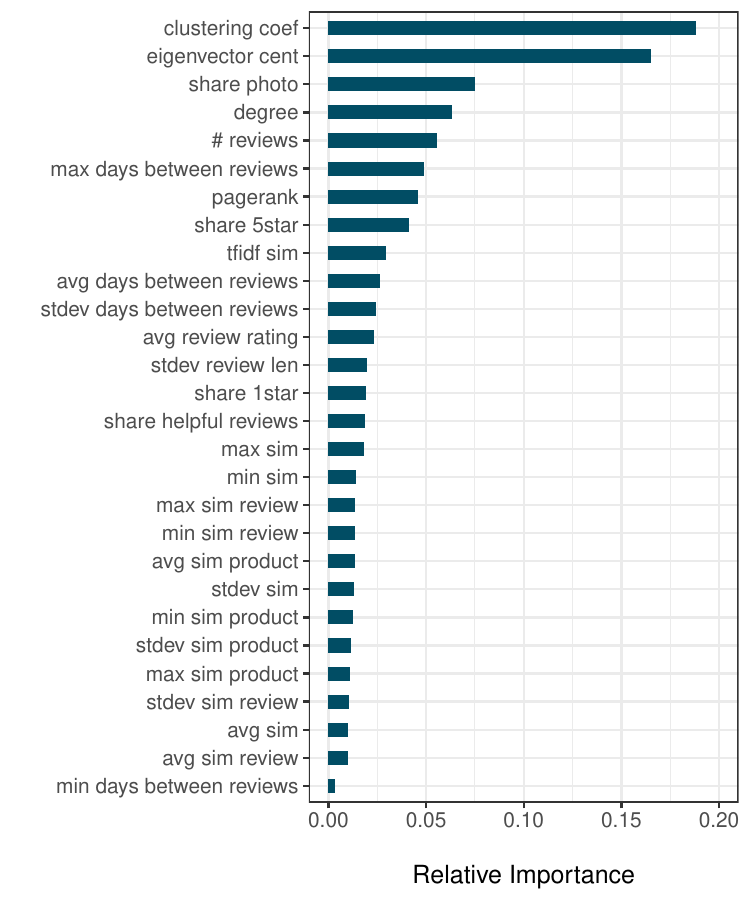}
    \caption{Relative importance of features in terms of their contribution to prediction performance of the random forests classifier including all features.}
    \label{fig:feature_imp}
\end{figure}

\begin{table}[ht]
\centering
\caption{Out-of-sample prediction performance of the random forests classifier with varying sets of features.}\label{tab:RF_pred_results}
\begin{tabular}{lccccc}
    \hline
    Features&	AUC&Accuracy	&TNR&	TPR&	F1 Score\\
    \hline
    \hline
    Network&\textbf{	.890} &	.\textbf{821} &	.839 &	\textbf{.797} &	\textbf{.821}\\
    Top-2 Network&	.879 &	.812 &	.832 &	.787 &	.812\\
    Metadata&	.874 &	.811 &	.858 &	.748 &	.810\\ 
    Text&	.857 &	.770 &	\textbf{.929} &	.559 &	.759\\
    Image&	.592 &	.599 &	.792 &	.343 &	.577\\
    \hline
    All Features&	.932 &	.860 &	.881 &	.832 &	.860\\ 
    \hline
\end{tabular}
\end{table}
Figure~\ref{fig:feature_imp} shows the relative importance of different individual features from the all-feature model in terms of their contribution to predictive power. The two most important features by a large margin are network features, specifically the clustering coefficient and the eigenvector centrality score. The fact that these two features rank highest in importance highlights the reason why the network-based model performs so well. Products buying fake reviews appear to rely on a common set of reviewers, causing them to be more closely clustered in the product network compared to regular products. While this result is novel in the scholarly literature, since platform techniques to detect fake reviews are not publicly disclosed, it may not be entirely novel in practice.

Table~\ref{tab:RF_pred_results} shows that when we test a more concise model using only these top two network features, it still outperforms all other models. This suggests that the predictive power of network features is sufficiently strong that, even when implemented in a simple way, they are more useful than models based on large sets of features derived from metadata, text, or images.

\subsection*{Unsupervised approach}
While we have shown that a model trained on features derived from the product network is highly accurate at detecting fake review buyers, platforms may not have the ground truth data required to estimate this type of model. Therefore, to further validate the strength of a network-based approach, we extend our analysis to a much larger dataset of Amazon product reviews where we lack ground truth data and test whether network data allow us to identify ex ante which products are likely to be manipulating their ratings. 

We use data that contains the entire universe of Amazon product reviews in the Home \& Kitchen category, which contains about 65 thousand products, 11 million reviews, and 6.1 million reviewers~\cite{ni2019justifying}. Because we no longer know which products buy fake reviews in this larger dataset, we follow an unsupervised approach.
We start by creating a product network as previously described. 
We then partition the products into 20 groups using the K-means clustering algorithm based on products' metadata and network features~\cite{raftery_clustering}. Once the products are clustered, we test if certain clusters are disproportionately likely to contain fake review products. To test this, we use our pre-trained classifier that was trained on all features to estimate the proportion of products buying fake reviews in each cluster. Although this calculation would ideally be done using ground-truth data, given the high out-of-sample accuracy of the all-features classifier, it should provide a good indication of the distribution of fake review products among clusters.

\begin{figure}[!ht]
\centering
\includegraphics[width=\linewidth]{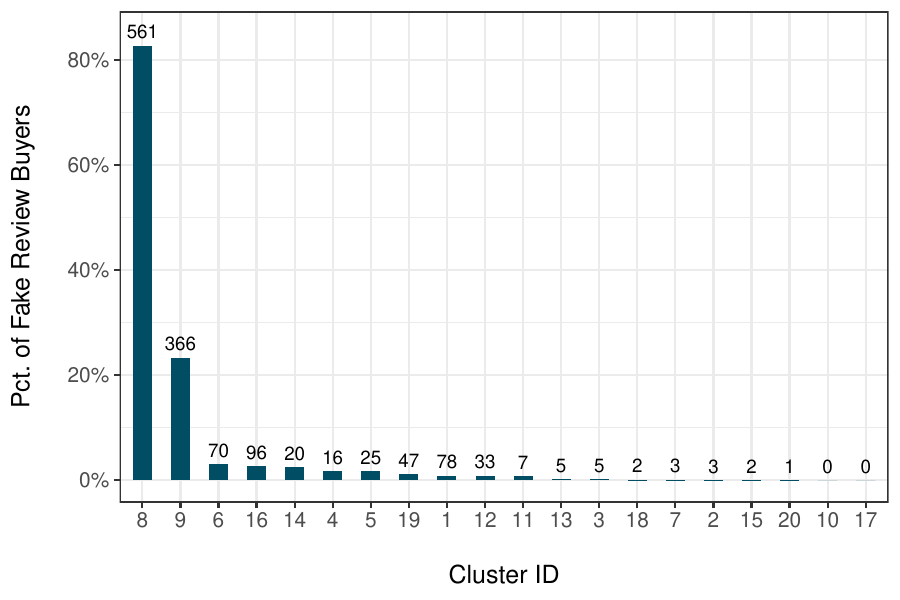}
\caption{The percentage of products that are identified as fake review buyers by the random forest classifier in each cluster.}\label{fig:clustering_results}
\end{figure}

Figure~\ref{fig:clustering_results} shows the percentage and the total number of products identified as buying fake reviews in each cluster using our classifier. We report summary statistics of clusters and the details of the unsupervised approach in Section 4 of SI Appendix.x	

This analysis shows that just two clusters (that contain only 3.4\% of products) contain about 70\% of the products identified as fake review buyers. In addition, within one of these clusters, a substantial majority of products are identified as buying fake reviews (83\%). In Table S4 (SI Appendix), we report the mean of the feature values in each cluster. The two clusters that contain the majority of fake review buyers are highly distinctive, and only in terms of their network features. This suggests that a platform could use unsupervised methods to identify these tightly-connected clusters of products without any ground truth data and use this information to identify likely fake review buyers.

\section*{Discussion and Conclusions}\label{sec:conclusion}
In this paper, we use a unique dataset and method to address a longstanding problem in the digital economy: rating manipulation. We use direct observation of what sellers buy fake reviews and propose a network-based approach to detect these products. 

Our analysis of the product network shows that products that buy fake reviews are highly clustered; while we don't have information about all Amazon reviews and reviewers and we only see reviews that are ultimately posted, this seems to suggests that reviewers participating in the fake review marketplace are a relatively small part of the full universe of Amazon reviewers. 
Therefore, a classifier based on just two network features, clustering coefficient and eigenvector centrality, can identify products buying fake reviews with high accuracy, outperforming models trained on large numbers of features constructed from metadata, review characteristics, text, and images. In addition to being powerful features to detect products that buy fake reviews, a crucial implication of this insight is that network-based features are very costly to manipulate because  these features result directly from the inherent limitations of acquiring reviews from fake review marketplaces. Although the specific setting we study is Amazon.com, the disproportionate clustering of rating manipulators in the network structure is likely to hold across applications where fake reviews are found. This is because it will generally be the case that the set of reviewers used by rating manipulators is likely to be substantially smaller than the general population of reviewers, leading to the same pattern of unusual clustering observed in our network. Moreover, since we show that no ground truth data is necessary to identify these clusters, platforms can easily implement our approach. This insight may also contribute to the study of similar problems, such as botnet-directed ad fraud~\cite{gordon_et_al_adfraud} and online review assessment~\cite{luan2018online}.

Interestingly, our results complement existing methods used by platforms to mitigate rating manipulation. In particular, platforms are often more cautious with new users with few reviews rather than with experienced reviewers with many reviews. While this approach is probably correct in general and is supported by prior research~\cite{luca2016fake}, our results suggest that there exist a set of experienced reviewers posting many fake reviews, and that by employing the network structure the platform can detect them.

Finally, while we provide a method that can detect products that buy fake reviews, we do not address the question of what to do with this information. Amazon can target products, sellers, reviewers, or all of them. For example, it can remove all product reviews, remove the product from the platform, ban the seller from selling on the platform, or ban or sue fake reviewers. Although Amazon currently targets the writers of fake reviews, in our view, Amazon should target sellers. 
An approach would be to ban these sellers. However, in cases where the false positive rate is non-zero this can be quite costly. Furthermore, it could provide a perverse incentive for sellers to buy fake reviews for their competitors to try to get them banned.
Therefore, a likely more balanced but still effective strategy would be to punish these sellers by reducing their visibility in the marketplace (e.g., by increasing their search rank) if they are suspected of buying fake reviews, which would offset and eliminate the financial incentive to doing so.

More generally, our results show that the product network structure provides a robust source of information that can allow rating platforms to apply greater scrutiny to the products they list which, in turn, can deter sellers from committing fraud. 

\acknow{We thank the Morrison Center for Marketing Analytics for generous funding. We thank seminar participants at the Johnson School of Management at Cornell University, the Ross School of Business at the University of Michigan, Pontificia Universidad Catolica de Chile, the Fox School of Business at Temple University, the RIT Marketing Workshop, the USC Marshall AI Workshop, and SCECR 2022 for helpful comments.}

\showacknow{} 

\bibliography{bibliography}

\begin{thebibliography}{10}

\bibitem{tadelis2016reputation}
S Tadelis, Reputation and feedback systems in online platform markets.
\newblock {\em\protect\JournalTitle{Annual Review of Economics}} \textbf{8},
  321--340 (2016).

\bibitem{levi_trust}
L Molm, R Hardin, M Levi, Cooperation without trust?
\newblock {\em\protect\JournalTitle{Administrative Science Quarterly}}
  \textbf{51}, 305--307 (2006).

\bibitem{he2022market}
S He, B Hollenbeck, D Proserpio, The market for fake reviews.
\newblock {\em\protect\JournalTitle{Marketing Science}} \textbf{41}, 896--921
  (2022).

\bibitem{Akesson_etal_welfare}
J Akesson, RW Hahn, RD Metcalfe, M Monti-Nussbaum, The impact of fake reviews
  on demand and welfare.
\newblock {\em\protect\JournalTitle{Working Paper}} (2022).

\bibitem{wu2020fake}
Y Wu, EW Ngai, P Wu, C Wu, Fake online reviews: Literature review, synthesis,
  and directions for future research.
\newblock {\em\protect\JournalTitle{Decision Support Systems}} \textbf{132},
  113280 (2020).

\bibitem{dong_trust}
B Dong, M Li, K Sivakumar, Online review characteristics and trust: A
  cross-country examination.
\newblock {\em\protect\JournalTitle{Decision Science}} \textbf{50}, 537--566
  (2019).

\bibitem{Heydari2015DetectionOR}
A Heydari, M ali Tavakoli, N Salim, Z Heydari, Detection of review spam: A
  survey.
\newblock {\em\protect\JournalTitle{Expert Syst. Appl.}} \textbf{42},
  3634--3642 (2015).

\bibitem{Mukherjee2013WhatYF}
A Mukherjee, V Venkataraman, B Liu, NS Glance, What yelp fake review filter
  might be doing? in {\em ICWSM}.
\newblock (2013).

\bibitem{Crawford2015SurveyOR}
M Crawford, TM Khoshgoftaar, JD Prusa, AN Richter, HA Najada, Survey of review
  spam detection using machine learning techniques.
\newblock {\em\protect\JournalTitle{Journal of Big Data}} \textbf{2}, 1--24
  (2015).

\bibitem{Vidanagama2019DeceptiveCR}
DU Vidanagama, TP Silva, AS Karunananda, Deceptive consumer review detection: a
  survey.
\newblock {\em\protect\JournalTitle{Artificial Intelligence Review}}
  \textbf{53}, 1323--1352 (2019).

\bibitem{burtch_hon_bapna_reviews}
G Burtch, Y Hong, R Bapna, V Griskevicius, {Stimulating Online Reviews by
  Combining Financial Incentives and Social Norms}.
\newblock {\em\protect\JournalTitle{Management Science}} \textbf{64},
  2065--2082 (2018).

\bibitem{oak2021fault}
R Oak, Z Shafiq, The fault in the stars: Understanding the underground market
  of amazon reviews.
\newblock {\em\protect\JournalTitle{arXiv preprint arXiv:2102.04217}} (2021).

\bibitem{he2016deep}
K He, X Zhang, S Ren, J Sun, Deep residual learning for image recognition in
  {\em Proceedings of the IEEE Conference on Computer Vision and Pattern
  Recognition}.
\newblock pp. 770--778 (2016).

\bibitem{france2012distance}
SL France, JD Carroll, H Xiong, Distance metrics for high dimensional nearest
  neighborhood recovery: Compression and normalization.
\newblock {\em\protect\JournalTitle{Information Sciences}} \textbf{184},
  92--110 (2012).

\bibitem{zhang2022can}
M Zhang, L Luo, Can consumer-posted photos serve as a leading indicator of
  restaurant survival? evidence from yelp.
\newblock {\em\protect\JournalTitle{Management Science}} (2022).

\bibitem{kumar2018detecting}
N Kumar, D Venugopal, L Qiu, S Kumar, Detecting review manipulation on online
  platforms with hierarchical supervised learning.
\newblock {\em\protect\JournalTitle{Journal of Management Information Systems}}
  \textbf{35}, 350--380 (2018).

\bibitem{ni2019justifying}
J Ni, J Li, J McAuley, Justifying recommendations using distantly-labeled
  reviews and fine-grained aspects in {\em Proceedings of the 2019 Conference
  on Empirical Methods in Natural Language Processing and the 9th International
  Joint Conference on Natural Language Processing (EMNLP-IJCNLP)}.
\newblock pp. 188--197 (2019).

\bibitem{raftery_clustering}
C Fraley, AE Raftery, {Model-based clustering, discriminant analysis, and
  density estimation}.
\newblock {\em\protect\JournalTitle{Journal of the American Statistical
  Association}} \textbf{97}, 611--631 (2006).

\bibitem{gordon_et_al_adfraud}
BR Gordon, et~al., {Inefficiencies in Digital Advertising Markets}.
\newblock {\em\protect\JournalTitle{Journal of Marketing}} (2021).

\bibitem{luan2018online}
S Luan, A Mueen, M Faloutsos, AJ Minnich, Online review assessment using
  multiple sources (2018) US Patent 10,089,660.

\bibitem{luca2016fake}
M Luca, G Zervas, Fake it till you make it: Reputation, competition, and yelp
  review fraud.
\newblock {\em\protect\JournalTitle{Management Science}} \textbf{62},
  3412--3427 (2016).

\end{thebibliography}

\end{document}



\maketitle

\section{Details of the features generation process}\label{apx:features}

\subsection*{Network features}\label{apx_sec:network_features}
In this section, we describe how we compute the network features. The degree of a product is calculated as follows:
\begin{equation}
    d_i = \sum_{j \in N_i} r_{ij},
\end{equation}
where $d_i$ is the degree of product $i$, $N_i$ is the set of products that have reviewers in common with product $i$, and $r_{ij}$ is the total number of reviewers that products $i$ and $j$ have in common. 

We calculate the eigenvector centrality of a product as follows:
\begin{equation}\label{equ:eigenvector_cent}
	e_i = \lambda_1^{-1}\sum_{j\in N_i}[\textbf{A}]_{ij}e_j,
\end{equation}
where \textbf{A} is the adjacency matrix of the network, $e_i$ is the eigenvector centrality of product $i$, and $\lambda_1$ is the largest eigenvalue of \textbf{A}. The $i,j$th element of \textbf{A} (i.e., $[\textbf{A}]_{ij}$) is 1 if products $i$ and $j$ have at least one reviewer in common, 0 otherwise. Notice that because the network is undirected with no self-loops, $[\textbf{A}]_{ij} = [\textbf{A}]_{ji}$ and $[\textbf{A}]_{ii} = 0$. With Equation (\ref{equ:eigenvector_cent}), a product's centrality score increases by having reviewers in common with other products that themselves share many reviewers with other products in the network. 

PageRank is an extension of eigenvector centrality~\citep{jackson2010social}, which we calculate as follows:
\begin{equation}\label{equ:pagerank}
    p_i = \frac{1-\alpha}{n} + \alpha\sum_{j\in N_i}[\textbf{A}]_{ij}\frac{p_j}{|N_i|},
\end{equation}
where $p_i$ is the PageRank of product $i$, $\alpha$ is the damping factor, $n$ is the total number of products in the network, and $|\cdot|$ is the cardinality of a set. PageRank mainly differs from eigenvector centrality by normalizing a neighbor product's structural importance by the number of its neighbors. 

We calculate the clustering coefficient of a product as follows:
\begin{equation}\label{equ:clustering_coef}
    c_i = \frac{2\sum_{j,k\in N_i}[\textbf{A}]_{ij}}{|N_i|(|N_i|-1)},
\end{equation}
where $c_i$ is the clustering coefficient of product $i$. The clustering coefficient takes values between 0 and 1.  

\subsection*{Computation of TF-IDF features}\label{apx_sec:tfidf}
We calculate the TF-IDF of a word in a product review as follows:
\begin{equation}\label{equ:tfidf}
    tfidf(w,r,R) = tf(w,r) + idf(w,R),
\end{equation}
where $w$ is the word, $r$ is the product review, $R$ is the set of a product's all reviews,
\begin{equation}\label{equ:tf}
    tf(w,r) = \frac{f_{w,r}}{\sum_{k\in r}f_{k,r}},
\end{equation}
and
\begin{equation}\label{equ:idf}
    idf(w,R) = \log\left(\frac{|R|}{|\{r\in R : w\in r\}|}\right).
\end{equation}
In Equation (\ref{equ:tf}), $f_{w,r}$ is the frequency of word $w$ in review $r$ and the denominator represents the total number of words in a review. According to Equation (\ref{equ:idf}), a word that frequently appears across a product's reviews will get a low score.

\subsection*{Image features}\label{apx_sec:image_features}
\begin{figure}[!ht]
\centering
\includegraphics[width=\textwidth]{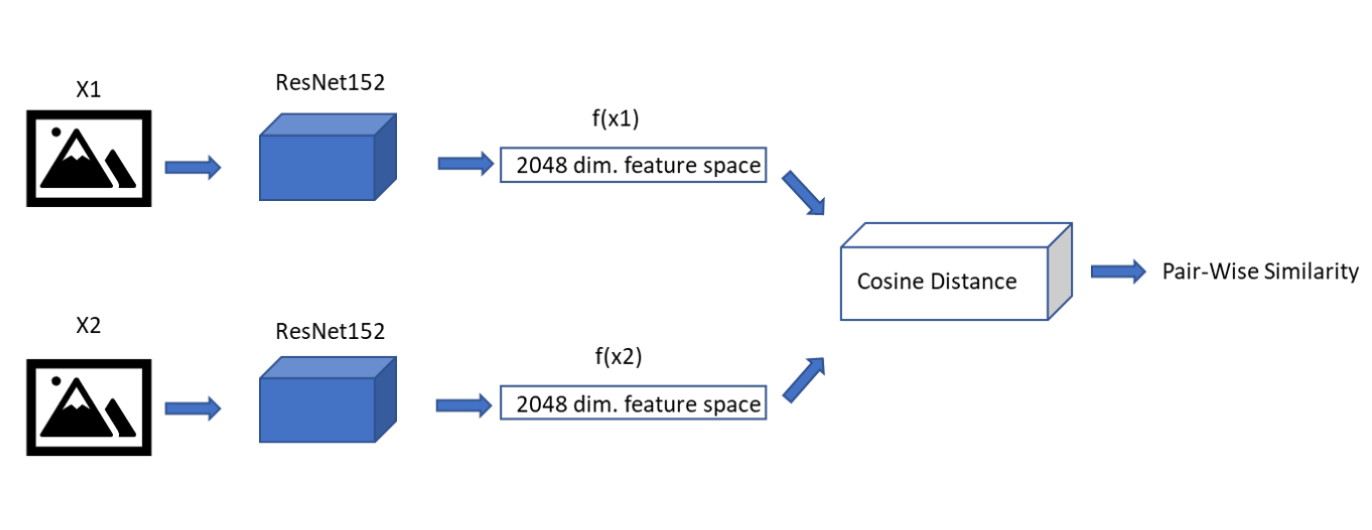}
\caption{Illustration of the process of calculating the pair-wise similarity between two images. 1) Extract a feature representation of each image using a pre-trained CNN (ResNet152). 2) Calculate the cosine distance between the two feature representations to obtain the similarity. The shorter the distance the higher the similarity. }
    \label{fig:im_sim}
\end{figure}

Figure~\ref{fig:im_sim} visualizes the process of calculating the similarity between pairs of images. First, we extract features using the ResNet-152, a deep Convolutional Neural Network (CNN) pre-trained on the ImageNet database~\citep{he2016deep}. Next, we use the output of the last fully connected layer to get a 2048 dimensional vector representation of each image. This layer roughly represents information related to objects, patterns, and colors. It structurally processes each image the same way, which allows us to compare them.

\subsection*{Correlation matrix}\label{apx_sec:tfidf}
In Table~\ref{tab:corr_mat}, we report the pairwise correlation coefficients between all features.

\begin{sidewaystable}[ht]
\centering
\caption{Correlation coefficients between features.}\label{tab:corr_mat}
\resizebox{\textwidth}{!}{\begin{tabular}{l*{27}{|c}}
\hline
&1&2&3&4&5&6&7&8&9&10&11&12&13&14&15&16&17&18&19&20&21&22&23&24&25&26&27\\
\hline
1. tfidf sim&-&&&&&&&&&&&&&&&&&&&&&&&&&&\\
2. \# reviews&0.24&-&&&&&&&&&&&&&&&&&&&&&&&&&\\
3. avg review rating&0.28&0.11&-&&&&&&&&&&&&&&&&&&&&&&&&\\
4. avg days between reviews&-0.14&-0.15&-0.02&-&&&&&&&&&&&&&&&&&&&&&&&\\
5. stdev days between reviews&-0.13&-0.19&-0.01&0.80&-&&&&&&&&&&&&&&&&&&&&&&\\
6. max days between reviews&-0.12&-0.08&-0.01&0.48&0.73&-&&&&&&&&&&&&&&&&&&&&&\\
7. min days between reviews&-0.09&-0.04&-0.02&0.76&0.27&0.19&-&&&&&&&&&&&&&&&&&&&&\\
8. share helpful&-0.16&-0.21&-0.16&0.16&0.19&0.14&0.06&-&&&&&&&&&&&&&&&&&&&\\
9. share 1star&-0.24&-0.11&-0.94&-0.02&-0.02&-0.03&-0.02&0.13&-&&&&&&&&&&&&&&&&&&\\
10. share 5star&0.28&0.08&0.94&-0.06&-0.05&-0.06&-0.05&-0.17&-0.80&-&&&&&&&&&&&&&&&&&\\
11. share photo&0.20&-0.11&0.18&-0.06&-0.09&-0.14&-0.01&-0.04&-0.12&0.24&-&&&&&&&&&&&&&&&&\\
12. std review len&0.32&0.07&0.00&-0.10&-0.08&-0.03&-0.06&0.17&-0.03&-0.03&0.19&-&&&&&&&&&&&&&&&\\
13. pagerank&0.40&0.56&0.15&-0.17&-0.22&-0.18&-0.05&-0.25&-0.10&0.19&0.16&0.09&-&&&&&&&&&&&&&&\\
14. degree&0.39&0.45&0.15&-0.16&-0.20&-0.18&-0.04&-0.24&-0.09&0.19&0.20&0.08&0.99&-&&&&&&&&&&&&&\\
15. clustering coef&0.08&-0.20&0.07&-0.10&-0.09&-0.13&-0.06&-0.03&-0.02&0.13&0.28&0.07&0.05&0.10&-&&&&&&&&&&&&\\
16. eigenvector cent&0.41&0.30&0.15&-0.18&-0.23&-0.22&-0.05&-0.25&-0.07&0.22&0.27&0.08&0.89&0.92&0.28&-&&&&&&&&&&&\\
17. min sim&0.14&-0.05&0.08&-0.24&-0.21&-0.10&-0.15&-0.01&-0.09&0.09&0.19&0.06&0.00&0.01&0.13&0.05&-&&&&&&&&&&\\
18. max sim&0.41&0.27&0.16&-0.36&-0.37&-0.20&-0.17&-0.15&-0.15&0.16&0.29&0.19&0.30&0.27&0.06&0.30&0.77&-&&&&&&&&&\\
19. avg sim&0.33&0.17&0.14&-0.33&-0.32&-0.17&-0.17&-0.11&-0.14&0.15&0.27&0.14&0.21&0.20&0.09&0.23&0.90&0.96&-&&&&&&&&\\
20. stdev sim&0.37&0.24&0.11&-0.29&-0.31&-0.16&-0.13&-0.12&-0.10&0.12&0.21&0.18&0.26&0.23&0.05&0.27&0.36&0.79&0.65&-&&&&&&&\\
21. min sim review&0.18&-0.04&0.10&-0.25&-0.22&-0.11&-0.15&-0.02&-0.10&0.11&0.23&0.09&0.03&0.03&0.14&0.08&0.98&0.80&0.92&0.42&-&&&&&&\\
22. max sim review&0.41&0.29&0.16&-0.37&-0.37&-0.20&-0.17&-0.16&-0.16&0.16&0.29&0.17&0.31&0.28&0.05&0.30&0.77&0.99&0.95&0.77&0.79&-&&&&&\\
23. avg sim review&0.33&0.17&0.14&-0.33&-0.32&-0.17&-0.17&-0.11&-0.14&0.15&0.28&0.14&0.21&0.20&0.09&0.23&0.90&0.96&1.00&0.66&0.92&0.96&-&&&&\\
24. stdev sim review&0.34&0.27&0.09&-0.28&-0.30&-0.14&-0.12&-0.13&-0.09&0.09&0.16&0.14&0.23&0.20&0.01&0.22&0.30&0.71&0.57&0.91&0.31&0.73&0.58&-&&&\\
25. min sim product&0.21&0.03&0.10&-0.29&-0.26&-0.14&-0.16&-0.04&-0.10&0.10&0.23&0.11&0.07&0.07&0.12&0.11&0.95&0.85&0.94&0.50&0.95&0.84&0.93&0.43&-&&\\
26. max sim product&0.39&0.27&0.15&-0.36&-0.36&-0.19&-0.17&-0.14&-0.15&0.15&0.27&0.19&0.30&0.27&0.07&0.30&0.79&0.97&0.95&0.74&0.81&0.96&0.95&0.67&0.87&-&\\
27. avg sim product&0.34&0.18&0.14&-0.34&-0.34&-0.18&-0.17&-0.11&-0.13&0.15&0.27&0.16&0.22&0.21&0.10&0.24&0.87&0.95&0.98&0.68&0.89&0.94&0.98&0.60&0.94&0.97&-\\
28. stdev sim product&0.36&0.26&0.12&-0.28&-0.29&-0.14&-0.13&-0.12&-0.12&0.12&0.15&0.18&0.26&0.23&0.01&0.25&0.42&0.73&0.64&0.78&0.45&0.72&0.64&0.71&0.47&0.78&0.68\\
\hline
\end{tabular}}
\end{sidewaystable}

\clearpage

\section{Details of the classification model}\label{apx:supervised_approach}
A random forests classifier is an ensemble model that uses a multitude of decision trees as base learners. A decision tree recursively splits the input space into nonoverlapping regions by minimizing the total variance across classes. Each tree is built on bootstrapped training observations and randomly selected input features and assigns an observation to the most commonly occurring class in the region it belongs. Then, the class of an observation is decided by the majority vote of trees. Below, we detail the model building process.

\subsection*{Parameter tuning} Random forests classifier has some parameters that need to be tuned. Table~\ref{tab:parameters} shows the tuned parameters and the values we use in our analysis. Before we test the model's performance, we tune these parameters. We start with random search cross-validation (CV) approach. Random search CV allows us to try a wide range of parameter values without computing all the combinations. In this step, we try [200, 400, 600, 800, 1000, 1200, 1400, 1600, 1800, 2000] for n\_estimators, [1, 2, 4] for min\_samples\_leaf, [2, 5, 10] for min\_samples\_split, ['auto', 'sqrt'] for max\_features, [10, 20, 30, 40, 50, 60, 70, 80, 90, 100, None] for max\_depth, and [True, False] for bootstrap. After 100 iterations of random search with 5-fold CV (i.e., total 500 different parameter combinations), the model performed the best with n\_estimator=200, min\_samples\_leaf=4, min\_samples\_split=2, max\_features=auto, max\_depth=20, and bootstrap=True. Using the results of the random search CV, we narrow the range of parameter values and follow the grid search approach with a 5-fold CV. Unlike random search CV, grid search evaluates all parameter combinations we define. In this step, we try [100, 200, 300] for n\_estimators, [3, 4, 5] for min\_samples\_leaf, [2, 4, 6] for min\_samples\_split, and [10, 20, 30, 40] for max\_depth. After evaluating all combinations, the values we use are shown in Table~\ref{tab:parameters}.
\begin{table}[ht]
\centering
\small
\caption{Hyperparameters of random forests classifier.}\label{tab:parameters}
\begin{tabular}{c|l|c}
\hline
Hyperparameter & Description & Value used \\
\hline
n\_estimators        & The number of decision trees to use as base learners & 100\\
min\_samples\_leaf    & The min. number of data points required in a leaf node & 3\\
min\_samples\_split   & The min. number of data points required before splitting a node & 6\\
max\_features         & The max. number of features considered to split a node & auto\\
max\_depth           & The max. number of levels allowed in a decision tree & 40\\
bootstrap           & Whether to use bootstrapped samples when building trees & True\\
\hline
\end{tabular}
\end{table}

\subsection*{Performance measures} To evaluate the model's performance, we train it using the randomly selected 80\% of the products and test it on the remaining 20\%. Before training the model, we standardize the features by removing the mean and scaling it to the unit variance to avoid the effect of varying scales of features in the dataset. We then report the model's AUC, accuracy, and F1 score on the test set. The receiver operating characteristic (ROC) curve shows the prediction performance of a classifier for all classification thresholds. It plots two values: true-positive rate (TPR) and false-positive rate (FPR). TPR (also known as recall or sensitivity) is the ratio of true positive observations to actual positive observations, and FPR is the ratio of true negative observations to actual negative observations. AUC measures the total area under the ROC curve. Accuracy is the percentage of observations that are correctly classified. F1 score is the harmonic mean of precision and recall, where precision is the ratio of true positive observations to all positive predicted observations. The higher the AUC, Accuracy, and F1 score values, the better the classification performance.


\section{Additional classifiers}\label{apx:additional_models}

To further validate our analysis, we test other classification models and obtain qualitatively similar results. The additional models include logistic regression, support vector machine (SVM), and extreme gradient boosting (XGBoost). 

A logistic regression classifier models the relationship between the probability that a seller is a fake review buyer and input features using a logistic function. The goal is to estimate the coefficients that are associated with input features. We use maximum likelihood to estimate the coefficients of the model. The SVM classifier obtains a $(p-1)$ dimensional hyperplane that separates the classes of training observations, where $p$ is the total number of features in the dataset. We use linear SVM with a regularization parameter of 1.0. XGBoost, similar to random forests, is also an ensemble learning model that uses decision trees as base learners to make a prediction. XGBoost uses a gradient descent algorithm to train the model. Like random forests, the XGBoost classifier has hyperparameters that need to be provided, such as the number of decision trees, the learning rate, the maximum depth for each tree, and regularization parameters. We use the default parameter values provided by the Python library of XGBoost in our analysis.

Table~\ref{tab:additional_results} shows the prediction performance of different classifiers with different sets of information. Results are aligned with the results of random forests classifier.

\captionsetup{width=0.6\textwidth}

\begin{table}[ht]
\centering
\caption{Comparisons of fake review buyer detection performance of additional classifiers with varying sets of features.}\label{tab:additional_results}
\begin{tabular}{c|l|cccccc}
\hline
\multirow{2}{*}{Model} & \multirow{2}{*}{Metric} & \multicolumn{6}{c}{Features} \\
\cline{3-8}
& & Image & Text & Metadata & Top-2 Network & Network & All Features \\
\hline
\multirow{5}{*}{\rotatebox[origin=c]{90}{Log. Reg.}} 
& AUC & 0.631  & 0.769  & 0.838 &  0.855  & 0.871  & 0.921 \\
& Acc & 0.620  & 0.731  & 0.785 &  0.793  & 0.802  & 0.857 \\
& TNR & 0.842  & 0.774  & 0.845 &  0.874  & 0.887  & 0.900 \\
& TPR & 0.325  & 0.675  & 0.706 &  0.685  & 0.689  & 0.800 \\
& F1  & 0.591  & 0.731  & 0.784 &  0.790  & 0.799  & 0.857 \\
\hline
\multirow{5}{*}{\rotatebox[origin=c]{90}{SVM}} 
& AUC & 0.629  & 0.749  & 0.839 &  0.854  & 0.874  & 0.920 \\
& Acc & 0.593  & 0.707  & 0.787 &  0.799  & 0.796  & 0.854 \\
& TNR & 0.900  & 0.750  & 0.858 &  0.855  & 0.911  & 0.892 \\
& TPR & 0.185  & 0.650  & 0.692 &  0.724  & 0.643  & 0.804 \\
& F1  & 0.529  & 0.707  & 0.785 &  0.797  & 0.790  & 0.854 \\
\hline
\multirow{5}{*}{\rotatebox[origin=c]{90}{XGBoost}} 
& AUC & 0.578  & 0.856  & 0.868 &  0.873  & 0.881  & 0.937 \\
& Acc & 0.571  & 0.773  & 0.800 &  0.809  & 0.817  & 0.877 \\
& TNR & 0.726  & 0.863  & 0.826 &  0.821  & 0.834  & 0.903 \\
& TPR & 0.364  & 0.654  & 0.766 &  0.794  & 0.794  & 0.843 \\
& F1  & 0.557  & 0.770  & 0.800 &  0.810  & 0.817  & 0.877 \\
\hline
\end{tabular}
\end{table}


\section{Details of the unsupervised approach}\label{apx:unsupervised_approach}

To apply the insights from our analysis on the smaller Amazon dataset to the larger one, we first create a product network. The network has about 65,000 nodes (i.e., products) and 16.2 million edges (i.e., existence of shared reviewers). Using the product network and review content, we then obtain the network features (degree, pagerank, eigenvector centrality, and clustering coefficient) and metadata features of the products in the larger dataset.

After obtaining the network and metadata features, we partition the products into 20 clusters using the K-means clustering algorithm. K-means is an unsupervised machine learning model that does not require ground truth labels and partitions the observations in a dataset into $K$ distinct and non-overlapping clusters. To do so, the K-means clustering algorithm minimizes the total within-cluster variation over all clusters. Within-cluster variation of a cluster is the average of squared distances between pairs of observations in the cluster. The steps of the K-means clustering algorithm are summarized as follows:
\begin{enumerate}
    \item Randomly assign products into one of $K$ clusters.
    \item Iterate until there is no changes.
    \begin{enumerate}
        \item Calculate the centroids of all clusters.
        \item Assign the products to the cluster whose centroid is the closest.
    \end{enumerate}
\end{enumerate}
We use Euclidean distance to measure the distances between pairs of observations in our analysis. To avoid the effect of varying scales of features, we standardize them by removing the mean and scaling to unit variance.

In Table~\ref{tab:clust_feats}, we report the mean of the feature values in each cluster.


\captionsetup{width=1.1\textwidth}

\begin{sidewaystable}[ht]
\centering
\small
\caption{Average of products' feature values in each cluster.}\label{tab:clust_feats}
\begin{tabular}{*{17}{c}}
\hline
\specialcell{Cluster\\ID} & \specialcell{Clust.\\Coef.} & \specialcell{Eig.\\Cent.} & \specialcell{Share\\Photo} & Degree & \# Rev. & \specialcell{Max\\Days} &\specialcell{Page\\Rank} & \specialcell{Share\\5star} & \specialcell{Tfidf\\Sim.} & \specialcell{Avg.\\Days}& \specialcell{Std.\\Days}   & \specialcell{Avg.\\Rating} &\specialcell{Std.\\Rev. Len.} & \specialcell{Share\\1star} & \specialcell{Share\\Helpful} & \specialcell{Min\\Days}\\
\hline
8&\cellcolor{coloring}2.21&\cellcolor{coloring}3.57&-0.34&\cellcolor{coloring}3.22&2.86&-0.34&\cellcolor{coloring}2.40&0.61&0.05&-1.19&-0.77&0.63&-0.34&-0.62&-0.91&-0.37\\
9&\cellcolor{coloring}3.41&-0.36&0.15&-0.32&-0.64&-0.37&0.28&0.34&-0.30&-0.05&-0.21&0.38&-0.09&-0.39&-0.35&-0.31\\
6&-0.41&-0.77&-0.09&-0.57&-0.70&-0.33&-0.24&0.55&-0.97&0.55&-0.11&0.53&-0.66&-0.51&-0.37&2.44\\
16&-0.32&-0.35&4.23&-0.40&-0.38&-0.61&-0.63&0.67&-0.45&-0.66&-0.62&0.60&-0.52&-0.52&-0.19&-0.35\\
14&-0.38&0.75&-0.30&-0.55&-0.36&3.88&1.78&0.31&-0.29&1.74&3.39&0.28&-0.33&-0.26&-0.15&-0.18\\
4&0.05&0.43&-0.83&-0.55&-0.60&1.05&1.43&-0.80&2.68&1.88&1.51&-0.62&1.84&0.51&2.73&-0.17\\
5&0.67&0.68&-0.31&0.06&-0.09&-0.36&0.39&-0.59&2.68&-0.39&-0.38&-0.44&3.51&0.35&1.63&-0.33\\
19&-0.29&-0.72&0.04&-0.39&-0.33&-0.37&-0.68&0.57&0.90&-0.42&-0.39&0.60&0.77&-0.59&-0.12&-0.37\\
1&-0.42&-0.88&-0.20&-0.45&-0.59&-0.48&-1.05&1.49&-0.82&-0.22&-0.39&1.23&-0.88&-0.96&-0.85&-0.37\\
12&-0.28&-0.65&-0.51&-0.48&-0.55&0.70&-0.67&0.97&-0.43&1.01&0.84&0.89&-0.57&-0.79&-0.40&-0.37\\
11&-0.58&-1.26&-0.58&-0.58&-0.71&0.49&1.54&-0.27&-0.28&2.10&0.98&-0.17&-0.01&0.08&0.48&3.47\\
13&-0.51&-0.02&0.01&-0.53&-0.42&-0.43&-0.26&-0.06&0.17&-0.24&-0.37&0.00&0.35&-0.03&2.25&-0.35\\
3&0.20&0.86&-0.25&2.25&2.48&-0.44&-0.30&0.39&0.14&-1.18&-0.79&0.39&-0.19&-0.39&-0.71&-0.37\\
18&-0.51&0.09&-0.18&-0.56&-0.41&-0.20&0.42&-2.81&-0.32&0.00&-0.13&-3.23&-0.11&3.62&0.09&-0.18\\
7&-0.35&-0.15&-0.52&-0.54&-0.57&0.37&-0.33&-1.00&-0.54&0.89&0.57&-0.78&-0.31&0.50&-0.01&-0.37\\
2&-0.47&-0.40&0.18&-0.48&-0.43&-0.52&-0.76&-1.51&-0.96&-0.50&-0.51&-1.46&-0.50&1.26&-0.47&-0.36\\
15&-0.63&-0.50&-0.03&0.04&0.07&-0.60&-0.99&1.14&-0.18&-0.91&-0.72&0.99&-0.63&-0.82&-0.81&-0.37\\
20&-0.51&-0.79&-0.06&-0.40&-0.51&-0.50&-1.04&-0.10&-1.07&-0.38&-0.47&0.03&-0.75&-0.19&-0.59&-0.37\\
10&-0.53&0.36&-0.23&0.17&0.83&-0.47&-0.56&-0.81&-0.05&-0.99&-0.70&-0.68&-0.19&0.48&-0.54&-0.37\\
17&-0.35&0.11&-0.16&1.07&1.05&-0.46&-0.74&0.91&0.05&-1.03&-0.72&0.83&-0.39&-0.73&-0.72&-0.37\\
\hline
\end{tabular}
\caption*{\normalfont \textit{Note:} The values of clusters are standardized. Features are listed from left to right in the order of their importance in predicting fake review buyers according to the random forests classifier. Distinctive network features of clusters that contain the majority of fake review buyers are highlighted in green.}
\end{sidewaystable}

\clearpage

\bibliography{bibliography}